# THE Fe-LINE FEATURE IN THE X-RAY SPECTRUM OF SOLAR FLARES: FIRST RESULTS FROM SOXS MISSION

Rajmal Jain, Anil K. Pradhan[+], Vishal Joshi, K. J. Shah, Jayshree J. Trivedi,
S. L. Kayasth, Vishal M. Shah, and M. R. Deshpande

Physical Research Laboratory
(Dept. of Space, Govt. of India)
Navrangpura, Ahmedabad – 380 009, India

+ Dept. of Astronomy,
The Ohio State University, USA

Running title: Fe-Line Feature in X-ray Solar Flares

Contact person:
Dr. Rajmal Jain
Email ID: rajmal@prl.res.in

**Abstract**: We present the first results from the "Low Energy Detector" payload of the "Solar X-ray Spectrometer (SOXS)" mission, which was launched onboard the GSAT-2 Indian spacecraft on 08 May 2003 by the GSLV-D2 rocket to study solar flares. The SOXS Low Energy Detector (SLD) payload was designed, developed, and fabricated by the Physical Research Laboratory (PRL) in collaboration with the Space Application Centre (SAC), Ahmedabad and the ISRO Satellite Centre (ISAC), Bangalore of Indian Space Research Organization (ISRO). The SLD payload employs state-of-the-art, solid-state detectors *viz.* Si PIN and Cadmium-Zinc-Telluride (CZT) devices that operate at near room temperature (-20 °C). The energy ranges of the Si PIN and CZT detectors are 4 - 25 keV and 4 - 56 keV respectively. The Si PIN provides sub-keV energy resolution while the CZT reveals ~1.7 keV energy resolution throughout the energy range. The high sensitivity and sub-keV energy resolution of the Si PIN detector allows measuring the intensity, peak energy, and the equivalent width of the Fe-line complex at approximately 6.7 keV as a function of time in all ten M-class flares studied in this investigation. The peak energy ($E_p$) of the Fe-line feature varies between 6.4 and 6.7 keV with increasing in temperature from 9 to 58 MK. We found that the equivalent width ($w$) of the Fe-line feature increases exponentially with temperature up to 30 MK and then increases very slowly up to 40 Mk. It remains between 3.5 and 4 keV in the temperature range of 30 - 45 MK. We compare our measurements of $w$ with calculations made earlier by various investigators and propose that these measurements may improve theoretical models. We interpret the variation of both $E_p$ and $w$ with temperature as due to the changes in the ionization and recombination conditions in the plasma during the flare interval and as a consequence the contribution from different ionic emission lines also varies.

# 1. Introduction

The "Solar X-ray Spectrometer (SOXS)" mission (Jain *et al.,* 2000a, b, 2005) was launched onboard an Indian geostationary satellite namely GSAT-2 on 08 May 2003 by the GSLV-D2 rocket. The SOXS Low Energy Detector (SLD) mission (Jain *et al.*, 2000a, b, 2005) aims to study the high energy and temporal resolution X-ray spectra from solar flares employing solid state detectors *viz.* Silicon PIN detector for 4 - 25 keV (area 11.56 sq. mm); and Cadmium Zinc Telluride (CZT) detector for 4 - 56 keV energy range (area 25 sq. mm). Details of the SLD instrumentation, the operation of the detectors, temporal and spectral resolution, and the data format were presented earlier by Jain *et al.,* (2005). The SLD payload was designed and developed at the Physical Research Laboratory (PRL) in collaboration with the ISRO Satellite Centre (ISAC), Bangalore, and the Space Application Centre (SAC), Ahmedabad.

The solar corona emits many X-ray lines below 10 keV and in order to improve our understanding of the X-ray line emission characteristics, synoptic observations at energies below 10 keV are of the utmost importance, since they may reveal the temperature enhancement during flares of different magnitude. On the other hand, it has been shown by Jain *et al.* (2000a, b, 2005) that iron complex lines (Fe XXV, XXVI) at 6.7 keV and Fe/Ni complex lines at 8 keV appear only during solar flare activity, however, understanding of their emission characteristics requires extremely high spectral and temporal resolution observations. The high sensitivity and sub-keV energy resolution of Si PIN detector allows the intensity and mean energy of the Fe-line complex at approximately 6.7 keV to be measured as a function of time in all classes of flares.

This line complex is due mostly to the 1s-2p transitions in Li-like, He-like, and H-like iron, Fe XXIV, Fe XXV and Fe XXVI respectively, with associated satellite lines. Another weaker line complex at ~ 8 keV made up of emission from He-like nickel and more highly excited Fe XXV ions is also evident in the more intense flares (Phillips, 2004, Phillips *et. al.*, 2004). Detailed calculations of emission line intensities as a function of temperature, with provision for different element abundance sets (*e.g.*, photospheric or coronal), are given by the MEKAL/SPEX atomic codes (Mewe, Gronenschild, and van den Oord, 1985; Mewe *et al.*, 1985, Phillips *et. al.*, 2004) and the CHIANTI code (Dere *et al.*, 1997). These codes also include thermal continuum intensities. They are used to interpret the SLD spectral observations in terms of the plasma temperature and emission measure. The centroid energy and width of the iron-line complex at ~6.7 keV, the intensity of the Fe/Ni line complex at ~ 8 keV, and the line-to-continuum ratio are functions of the plasma temperature and can be used to limit the range of possible plasma parameters. However detailed study of such features of the Fe and Fe/Ni line complexes has not been carried out earlier mainly due to the lack of spectral observations in the energy range 3 - 10 keV with high spectral and temporal resolution, which are critically required to measure precisely the line features and plasma parameters. The high spectral and temporal resolution spectra may reveal many unidentified lines as shown by the RESIK Bragg crystal spectrometer aboard CORONAS-F (Sylwester *et al.*, 2004). Phillips *et al.*, (2004) carried out a study of solar flare thermal spectra using RHESSI, RESIK, and GOES data and determined absolute elemental abundances, which however may be subject to uncertainties due to measurements from three different instruments that were not calibrated by a single common technique. However, the SOXS mission provides the X-ray spectra in the desired 4 - 10 keV energy band with improved spectral and temporal resolution. Therefore the purpose of this paper is to study the X-ray emission characteristics of the Fe-line feature in solar flares using the high sensitivity and sub-keV energy resolution capabilities of the Si PIN detector on the SOXS mission. We present the current study of the Fe-line emission as the first results from the observations made by the SLD/SOXS mission. In Section 2 we present the observations made by the SLD payload. Section 3 describes the analysis techniques and the results obtained. We discuss our findings in Section 4 and present conclusions in Section 5.

## 2. Observations

2.1 INSTRUMENTATION:

The instrumentation of the SLD payload, its in-flight calibration, and operation have been described by Jain *et al.,* (2005). However, a brief description of the experiment is given here. The SOXS consists of two independent payloads *viz.* SOXS Low Energy Detector (SLD) and SOXS High Energy Detector (SHD). The SLD payload is functioning satisfactorily onboard the GSAT-2 spacecraft and so far more than 300 flares of importance greater than GOES C1.0 have been observed. The spectral resolution of the Si detector is 0.7 keV at 6 keV and 0.8 keV at 22.2 keV, which is better than that of earlier detectors used for solar flare research in this energy range. However, the spectral resolution achieved from CZT detector is poorer *i.e.,* almost 1.7 keV, but it remains stable throughout its energy range of 4 – 56 keV. Further the temporal resolution capabilities are also superb as we designed for 100 ms during flare mode in order to achieve statistically significant energy spectra.

We used an 8-bit ADC as a pulse-height analyzer for the Si and CZT detectors to form the spectra in the dynamic energy ranges 4 - 25 and 4 - 56 keV respectively. These show 0.082 keV and 0.218 keV channel widths for the Si and CZT detectors. Pre-launch, the Si and CZT detectors were characterized and calibrated (Jain *et al.,* 2003) using radioactive sources *viz.* $Fe^{55}$ emitting a line at energy 5.9 keV; $Cd^{109}$ at 22.2 and 25.0 keV, and $Am^{241}$ at 13.9, 17.8, 20.8, 26.3, 33.0, and 59.5 keV. The energy scale was established by setting the gain of the amplifier to match the observed peak energy of the $Fe^{55}$ standard radioactive source to its theoretically-known peak energy. In order to cross check the full energy scale (0 - 255 channels) we carried out the above experiment with $Cd^{109}$ and $Am^{241}$ radioactive sources. Once the energy scale was established, the gain was fixed with a selected fixed value resistor. Next, the peak detector and shaping amplifier were biased so as to operate in a highly linear region. The linearity between the energy (pulse height) scale and peak energy was observed with an uncertainty of ± one channel. The linearity was tested using the above radioactive sources during vibration and thermal vacuum tests and at the launch pad, and no variation was observed. In-flight calibration is carried out by an onboard $Cd^{109}$ weak radioactive source, which is mounted inside the collimator (Jain *et al.,* 2005), and emits lines at 22.2 and 25 keV. Integrating over long periods, we have calibrated many times and found that the peak of the 22.2 keV line for the Si and both lines for the CZT detectors falls at the same channel as observed pre-launch with an uncertainty of ± one channel.

The critical operating temperature of both the detectors in the range -5 to -30 $^0$C is achieved using thermoelectric coolers that are coupled to the detectors. The detector package is mounted on a Sun Aspect System which keeps the Sun in the center of the detector field of view for an interval between 03:40 – 06:40 UT every day. However after 06:40 UT, the temperature of the detectors exceeds the capability of the thermoelectric cooler. The SLD data is of two types: temporal mode (light curves) and spectral mode.

2.2 DATA SET:

The first light from the Sun was observed by the detectors on 08 June 2003. The flare trigger threshold was intentionally kept higher so as to observe the signal clearly above the background. The temporal data, *i.e*., intensity (counts/s) as a function of time, is obtained in four energy bands *viz*. 6 - 7 keV (L1), 7 - 10 keV (L2), 10 - 20 keV (L3), and 4 - 25 keV (T) by the Si detector and in five energy bands by the CZT detector *viz*. 6 - 7 keV, 7 - 10 keV, 10 - 20 keV, 20 - 30 keV, and 30 - 56 keV. In Table I we show the flare events analyzed to study the X-ray spectral evolution of the Fe-line feature in the flare plasma. We selected a total of ten flares of GOES importance class M for the current study. Data from the Si detector are used to study the Fe-line feature.

2.2.1 *Temporal Mode*:

In Figure 1 we show the temporal-mode observations *i.e.* light curves of the 31 October 2004 flare, in four energy channels for the Si detector. The time resolution for temporal and spectral mode observations during quiet periods is one second and three seconds respectively, but during flares it is 100 ms for both temporal and spectral modes. The intensity (counts/s) of the light curve shown in Figure 1 is a 20-second moving average of the 100 ms observed data. It may be noted that the flare is composed of a slowly rising thermal phase followed by a superhot phase. The flare was also observed by the GOES spacecraft as shown in Table I.

**Table I**

SLD/SOXS Flare Events Considered for Investigation

| S. No. | Date | Time UT Begin | Time UT Peak | Time UT End | SOXS (Si) *Peak Int. | GOES Class | Active Region Location | Active Region NOAA |
|---|---|---|---|---|---|---|---|---|
| 1. | 30 Jul 2003 | 0407 | 0409 | 0428 | 2420 | M2.5 | N16 W55 | 10422 |
| 2. | 13 Nov 2003 | 0454 | 0501 | 0510 | 1292 | M1.6 | N04 E85 | 10501 |
| 3. | 19 Nov 2003 | 0358 | 0402 | 0419 | 1190 | M1.7 | N01 E06 | 10501 |
| 4. | 07 Jan 2004 | <0355 | 0400 | 0433 | 2243 | M4.5 | N02 E82 | 10537 |
| 5. | 25 Mar 2004 | 0429 | 0438 | 0507 | 1740 | M2.3 | N12 E82 | 10582 |
| 6. | 25 Apr 2004 | 0528 | 0536 | 0558 | 1918 | M2.2 | N13 E38 | 10599 |
| 7. | 14 Jul. 2004 | 0518 | 0523 | >0525 | 4093 | M6.2 | N12 W62 | 10646 |
| 8. | 14 Aug 2004 | 0537 | 0544 | >0604 | 7474 | M7.4 | S11 W28 | 10656 |
| 9. | 31 Oct 2004 | 0526 | 0531 | 0546 | 2062 | M2.3 | N13 W34 | 10691 |
| 10. | 25 Aug 2005 | 0436 | 0439 | 0452 | 4705 | M6.4 | N07 E78 | 10803 |

Note: > – After, < – Before, * - Peak intensity in 7 – 10 keV in counts/sec.

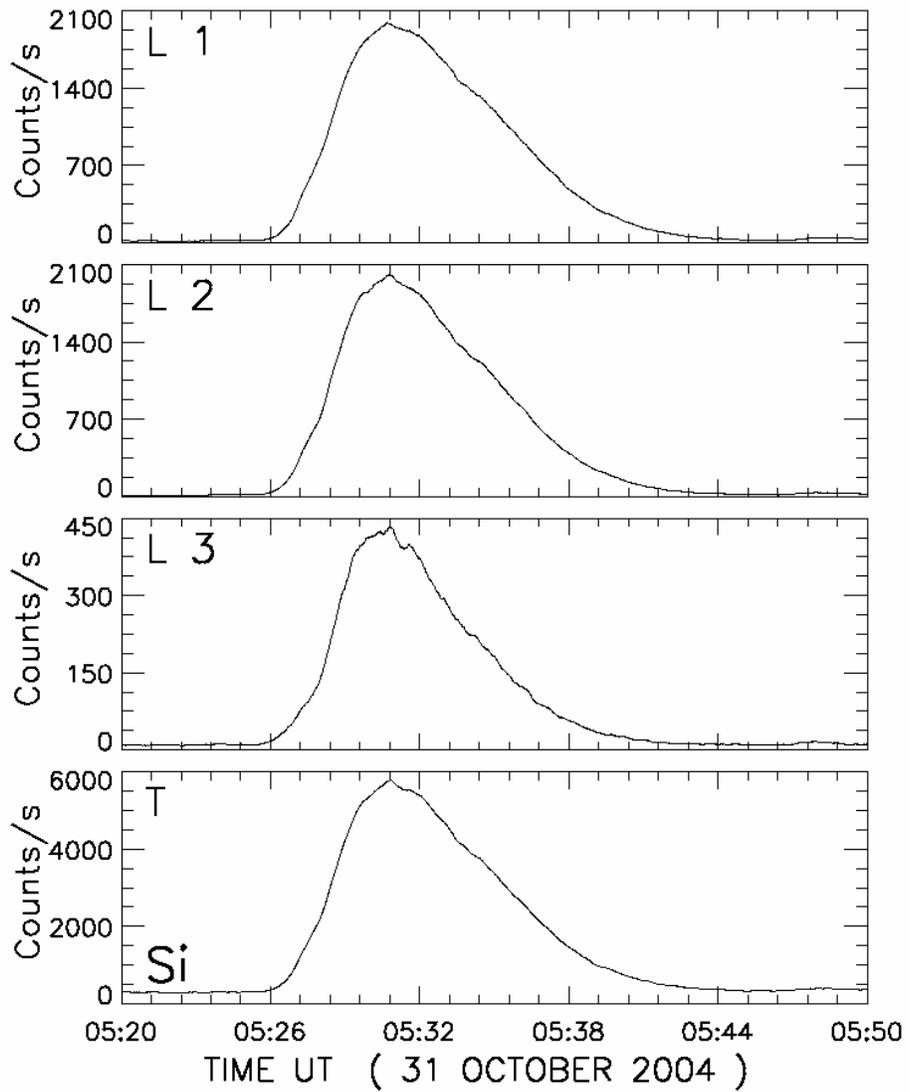

Figure 1: Light curves of the 31 October 2004 solar flare as recorded in L1, L2, L3, and T energy bands (see text) of the Si detector of the SLD/ SOXS mission

2.2.1 *Spectral Mode*:

The energy region 4 - 15 keV in solar flare X-ray spectra is of great importance for inferring the properties of the hottest parts of the thermal plasma created during a solar flare. It contains emission lines of highly ionized Ca, Fe, and Ni atoms and a continuum that falls off steeply with increasing energy. In this context SLD is the first payload which has an energy range of 4 - 25 keV to study the line emission and continuum with sub-keV spectral resolution. This is achieved by employing the Si PIN detector as described in the preceding section.

The energy spectrum, intensity (counts/s) as a function of energy at a given time, in the energy range 4 – 25 keV is distributed over 256 channels with a channel width of 0.082 keV. The detected count spectrum is in fact given by the convolution of the actual photon spectrum with the response matrix as shown in Equation (1). The response matrix of an X-ray detector allows reconstruction of the source photon spectra from the observed counts per channel in the Pulse Height Analyzer (PHA).

$$C_i = \int_0^{255} \frac{dN}{dE} R(i,E) dE = \sum_j \frac{dN}{dE}(E_j) R_{ij} \Delta E_j \qquad (1)$$

where $C_i$ are the detected counts in the *i*-th PHA channel, $dN/dE$ is the input photon spectrum, $R$ is the overall response matrix and *j* describes the binning of the photon input energy $E$, were the *j*-th bin has a width $\Delta E_j$. In Equation (1) the matrix $R_{ij}$ is an overall response matrix having dimension of [cm$^2$ keV], indicating the efficiency of the detector folded over effective area and FWHM (energy resolution).

The Si detector's count spectra at the peak time of the 31 October 2004 flare is shown in Figure 2. The low intensity below 6 keV is due to the aluminum and kapton filter mounted on the detector head to reduce the X-ray photon response below 4 keV and the electron response below 300 keV (Jain *et al.*, 2005). It may be noted that in the Si count spectra, the Fe and Fe/Ni lines are unambiguously visible at ~6.7 and ~8 keV respectively.

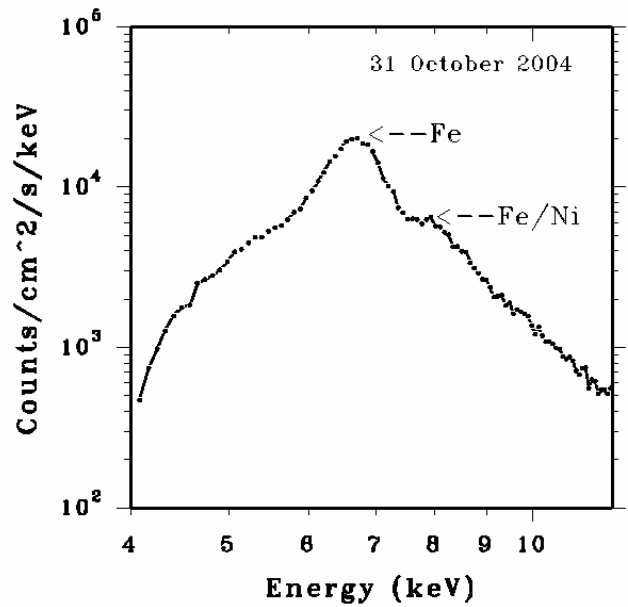

Figure 2: Count Spectra from the Si PIN detector for 31 October 2004 flare at 05:30:59 UT. Note Fe and Fe/Ni line features.

### 3. Analysis and Results

The raw data for temporal and spectral mode observations are first corrected for any spurious, or false, flares as well as for the pre-flare background (Jain *et al.*, 2005). The spectrum at a given time is formed by integrating the high cadence (100 ms) spectra over an interval of 30 to 100 seconds. The photon spectrum is produced by deconvolution of the count spectrum over the instrumental response as follows.

Let $N_{ij}$ be the corrected PHA spectral data where $i$ is a spectral record from 0 to $n$, and $j$ is the channel number ranging from 0 to 255 for that particular spectral record. Firstly, in order to calculate the background spectra we select a range of $N_{ij}$ where the Sun is quiet for a significant period (>20 minutes) between $i_b$ and $i_e$ on the observational interval. Here $i_b$ and $i_e$ are the beginning and ending spectral records for the quiet interval. Integrated background counts spectra ($IB_j$) may be written as follows.

$$IB_j = \frac{\sum_{i_b}^{i_e} N_{ij}}{T_{i_e} - T_{i_b}} \qquad (2)$$

Now, for generating a photon spectrum of the flare for a given interval *viz.* $k_b$ to $k_e$, during the flare, we first generate count spectra for this time interval as shown in relation (3).

$$IF_j = \frac{\sum_{k_b}^{k_e} N_{lj}}{T_{k_e} - T_{k_b}} \qquad (3)$$

However, to obtain pure flare count spectra ($CF_j$) we have to subtract the background count spectra ($IB_j$) from $IF_j$, so we obtain,

$$CF_j = (IF_j - IB_j) \qquad (4)$$

and finally the count spectra ($C_i$) are deconvolved over the instrumental response ($R_j$) to obtain the flare photon spectra ($PFj$) as shown in the relation (5).

$$PF_j = \frac{CF_j}{R_j} \qquad (5)$$

These photon spectra may be used to study the X-ray line and continuum emission. We demonstrate in Figure 3 that the flare signal is not contaminated with background by showing the photon spectra of the background and the flare observed on 31 October 2004. The various steps from data acquisition to data analysis were presented in detail by Jain *et al.,* (2005). The photon spectra used to study the evolution of Fe and Fe/Ni lines in a given flare are formed by integrating the spectra over 30 to 100 seconds depending on the observing cadence of 100 ms or three seconds during flare mode and quiet mode respectively.

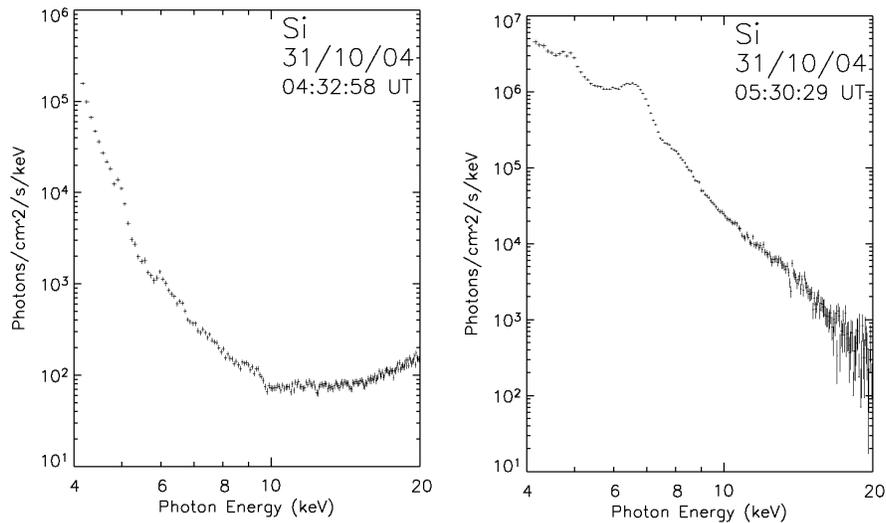

Figure 3: Photon spectra of background (left) and the flare (right) intervals on 31 October 2004. Note that photon flux in the flare is two - three orders higher than the background below 9 keV and one - two orders higher in the continuum.

3.1 X-RAY EMISSION FROM THE Fe-LINE:

In order to study the Fe and Fe/Ni line emission, it is important to study their evolution with the flare development *i.e.*, as a function of temperature, because the line emission and its intensity vary with temperature and emission measure (Phillips, 2004). Figure 4 shows a sequence of photon spectra for the 31 October 2004 flare in the energy range 5 - 12 keV. The sequence shows the evolution of the Fe and Fe/Ni lines as a function of time. It may be noted from this figure that the peak intensity, peak energy and area under the curve of the lines vary over time. In fact the plasma temperature and emission measure vary over time and these factors mainly control the shape of the line. Although the non-thermal contribution may also play a role; in this paper we consider only temperature and emission measure as important parameters.

The Fe line feature is defined here as the excess above the continuum, as observed by Si spectrometer with spectral resolution (FWHM) ≤ 0.7 keV, in the energy range 5.8 - 7.5 keV (Phillips, 2004). It may be noted from the temporal evolution of this line shown in Figure 4 that the Fe-line features including, peak energy, equivalent width (*w*) and intensity, vary during the flare. This suggests that the strength and peak energy of the emission line vary as a function of temperature. In this paper we investigate the variation in peak energy of the Fe-line feature as derived by a Gaussian fit, which allows us to measure the central peak energy and equivalent width (*w*) for a given spectra, as a function of the temperature. We analyzed 10 to 27 spectra, for each flare depending on the duration of the flare. The SOXSoft package (Patel and Jain, 2005) is used for data analysis.

Once the photon spectra are formed, we undertake their analysis for deriving plasma parameters such as temperature, emission measure, and spectral index using the SOXSoft spectra fitting program. This program takes its main routine from Solarsoft where the Mewe and Chianti codes can be used to derive the plasma parameters. In order to fit the spectra in the energy range between 5 and 15 keV, and particularly the Fe-line feature by isothermal plasma, we use the Chianti code because thermal continuum from it is within 1% of the detailed calculations of Culhane (1969) and the approximation of Mewe, Gronenschild, and van den Oord, (1985). We use the best-fit to the line feature based on the minimum reduced $\chi^2$ (difference counts). In order to derive the line parameters such as the peak energy ($E_p$), net and gross area under the curve, and the equivalent width, we subtracted the continuum contribution to the spectrum.

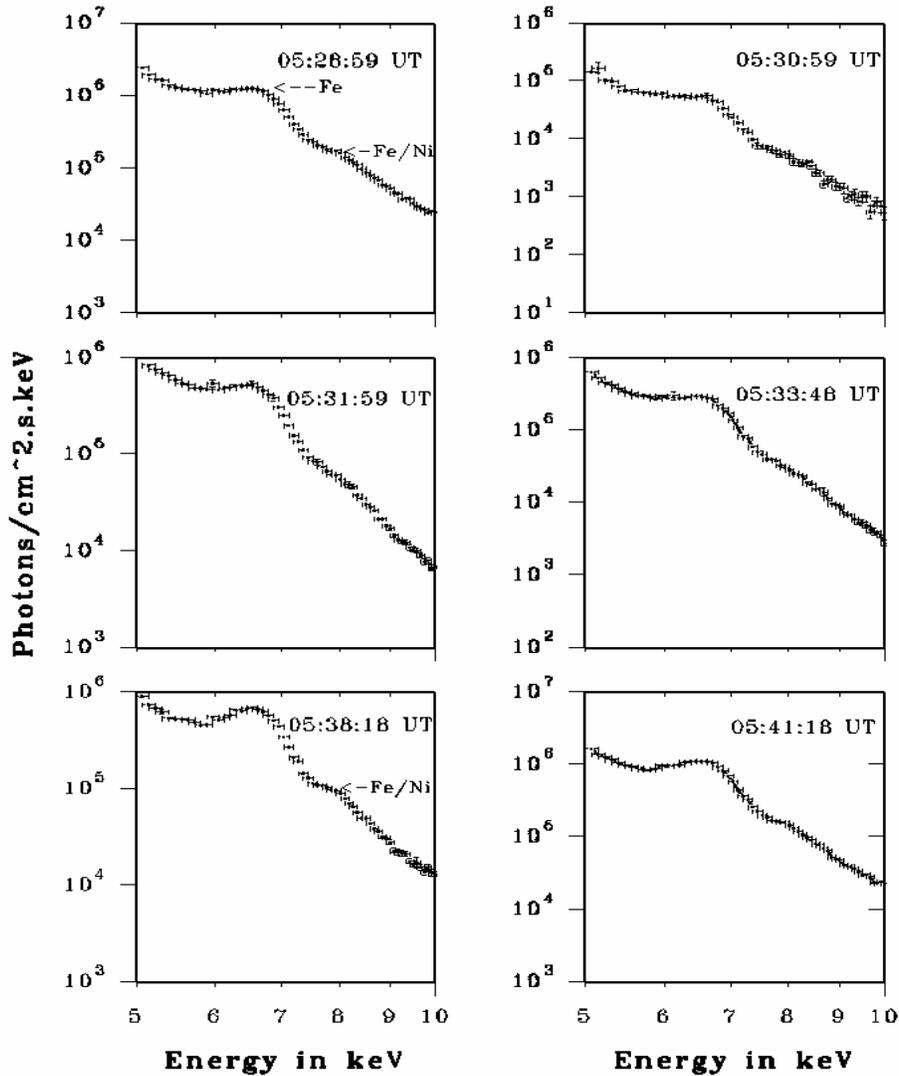

Figure 4: Sequence of *X*-ray photon spectra in the energy range 5 – 10 keV for the 31 October 2004 flare showing evolution of the Fe and Fe/Ni line features. The *X*-axis bar is the channel width of 0.082 keV. The *Y*-axis error bar is the ± 1σ value of the photon flux in the given channel.

3.1.1 *Evolution of Temperature and Emission Measure*

Temperatures are derived from the continuum part of the photon spectrum by a best-fit of photon flux from the Chianti code assuming on isothermal plasma temperature and emission measure for the energy range between 9.5 and 20 keV. In Figure 5, for example, we show best fits of photon flux for a single spectrum from each of the 14 August 2004 and 31 October 2004 flare events. The temperature and emission measure derived for these spectra are 57.4 MK and $1.5 \times 10^{49}$ cm$^{-3}$, and 29.3 MK and $1.3 \times 10^{49}$ cm$^{-3}$ respectively. The isothermal fit by the Chianti code using Solarsoft is accepted if the reduced $\chi^2 < 5$. For example the residual counts for the continuum fit for a $\chi^2$ of 0.59 and 1.55 (*cf.* Figure 5) are shown in Figure 6. In this way, we obtained temperature and emission-measure values for each photon spectrum at given times during the flare. We studied 10 to 27 integrated photon spectra for each flare depending upon its duration. In Figure 7, as an example, we show the temperature and emission measure evolution for the 14 August and 31 October 2004 flares. We compare the temperature evolution with the light curve for each individual flare event observed by the Si detector in the 4 - 25 keV range. We found that the evolution of the temperature is closely similar to the light curve of the flare, and peaks around flare maximum. This indicates that flare X-ray photon emission is rather strongly governed by the temperature of the plasma. On the other hand, the emission measure (*cf.* Figure 7) was between 1 and $2 \times 10^{49}$ cm$^{-3}$ in the 14 August 2004 flare and between 1 and $3 \times 10^{49}$ cm$^{-3}$ throughout the flare of 31 October 2004 except in one spectrum where it was seen higher than $4 \times 10^{49}$ cm$^{-3}$.

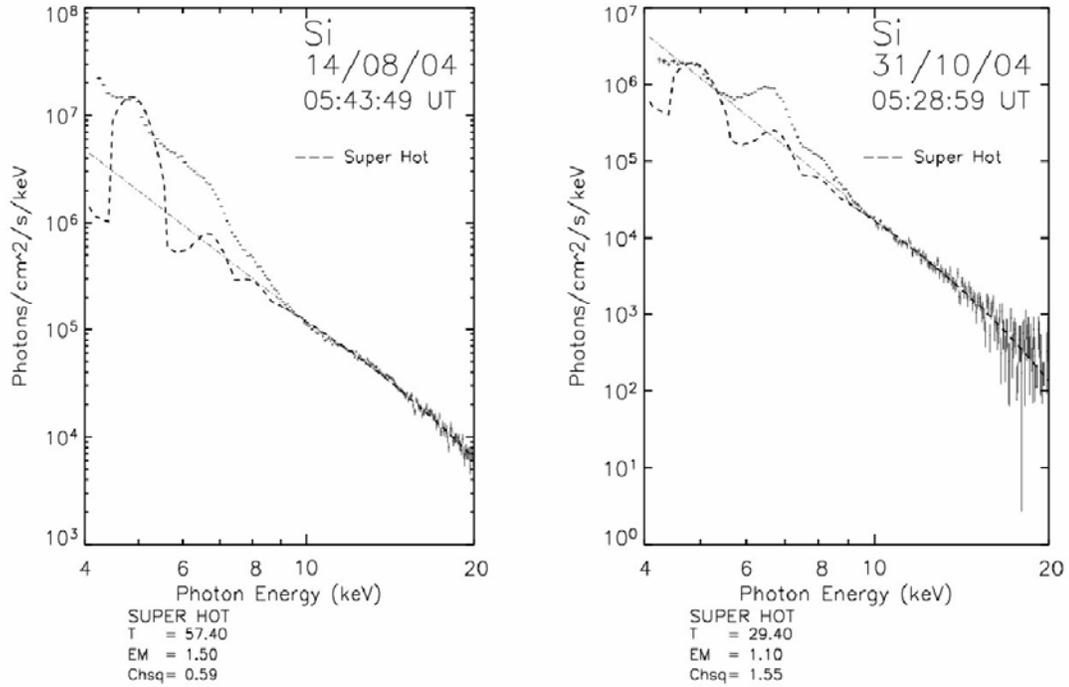

Figure 5: The X-ray photon spectra of the 14 August 2004 and 31 October 2004 at 05:43:49 UT and 05:28:59 UT respectively. Note the 9.5 – 16 keV continuum fit by isothermal plasma temperature (superhot) and emission measure. Contribution from the continuum emission down to 4 keV is shown by the extrapolated dash-dot straight line from the continuum fit. The *X*-axis bar is the channel width *i.e.* 0.082 keV, while the *Y*-axis error bar is ± 1σ of the photon flux in the given channel.

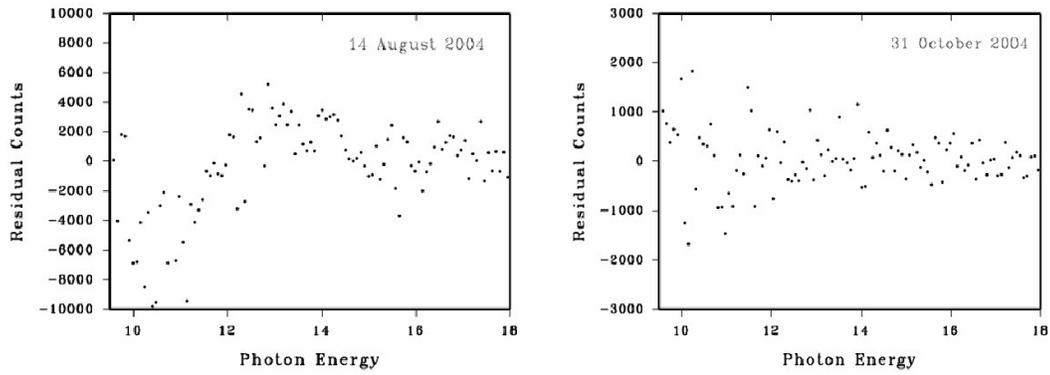

Figure 6: The residual counts for isothermal plasma continuum fit (*cf.* Figure 5) for the 14 August 2004 and 31 October 2004 spectra.

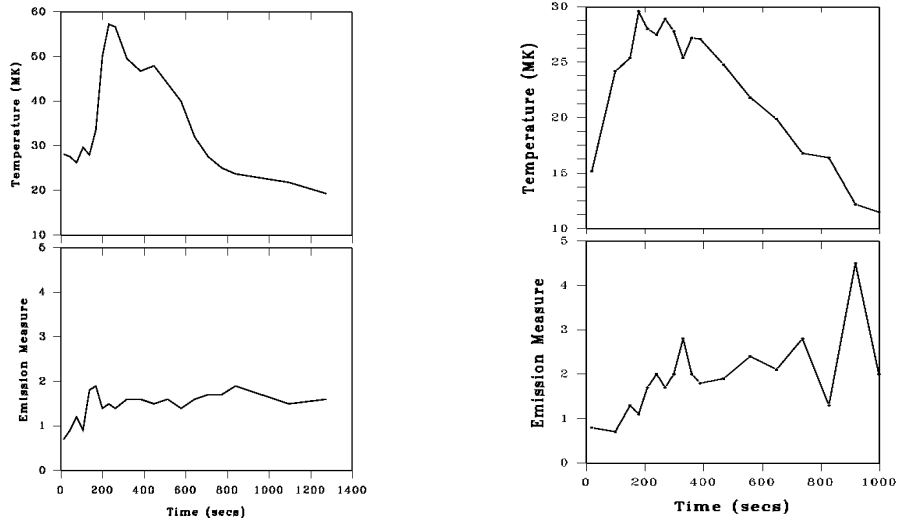

Figure 7: Evolution of temperature and emission measure derived from the continuum of the X-ray photon spectra of the 14 August 2004 (left) and 31 October 2004 (right) flares.

3.1.2 *Peak Energy of Fe-Line Features*

The thermal component in the Si PIN spectra is observed to have a prominent broadened emission line feature at 6.7 keV and a less intense line feature at 8 keV indicating high plasma temperatures. The 6.7 keV feature corresponds to a group of emission lines due to Fe XXV, associated dielectronic satellites of ions from Fe XIX to Fe XXIV, and fluorescence-formed lines of Fe II, and a second group due to Fe XXVI (Lyα) lines and associated satellites. The Fe XXV lines are excited at electron temperatures $T_e \geq 12$ MK, while the Fe XXVI lines are excited at $T_e \geq 30$ MK (Phillips, 2004). This line complex is referred to in this paper as the Fe-line feature. Thus we may conclude that the Fe-line feature is made up of many individual lines each having its own temperature dependence. Their contribution to the total emission of the Fe-line feature will therefore change as the temperature of solar flare plasma changes in both space and time. This results in changes to the energies of the Fe-line feature, as defined by the energy of the peak intensity ($E_p$). The variation in central peak energy of the Fe line feature may be between 6.3 and 7.0 depending on the temperature of the flare plasma that ranges from 10 to >100 MK. Changes in $E_p$, if large enough, therefore provide a possible useful temperature diagnostic. However, measurement of $E_p$ requires high spectral resolution, which the Si detector provides. In our case $E_p$ for the Fe line feature can be measured with an uncertainty of ± one channel *i.e.* ± 0.082 keV. Assuming that the large sample of spectra analyzed for ten flares may allow us to estimate the variation of $E_p$ for the Fe-line feature as a function of temperature, we derived the peak energy firstly fitting the line by Chianti codes and then by Gaussian fitting. The peak energy variation reveals the dominant physical process among H, He, and Li type ionic emissions, however including satellites and di-electronic emissions, of the Fe-line feature that taking place at a given temperature.

In Figure 8 we show the variation of peak energy ($E_p$) as a function of temperature for the Fe-line feature. We measured $E_p$ for each photon spectra in the flare for which the temperature had been derived from the continuum. A total of 135 spectra from all ten flares were analyzed to measure $E_p$. Later, in order to get better statistical confidence, we distributed the 135 $E_p$ values in 1 MK intervals according to their respective temperatures. For example, all $E_p$ measurements from the spectra falling in the temperature range between 9 and 11 MK were averaged and plotted at a mean $E_p$ value at 10 MK and also a standard deviation ($\sigma$) of the $E_p$ values was obtained. This mean $E_p$ value is shown with ± one $\sigma$ at a mean temperature of 10 MK. The $E_p$ value for each photon spectrum of each individual flare was derived using the line parameter software of Soxsoft that employs the Chianti code to derive plasma parameters as mentioned earlier. The *X*-axis bar is the channel width *i.e.* 0.082 keV corresponding to ~1 MK temperature uncertainty. It is the same throughout all the spectra analyzed by us, and therefore not represented in the plot. Figure 8 shows that $E_p$ increases with temperature in agreement with the results of Phillips (2004), and Oelgoetz and Pradhan (2004) but above 25 MK it decreases. Our results indicate that in the temperature range of 10 - 58 MK measured the Fe-line feature is mostly dominated by Fe XXV emission. However, this result is limited by the uncertainty of ± 0.082 keV in the measurements of $E_p$.

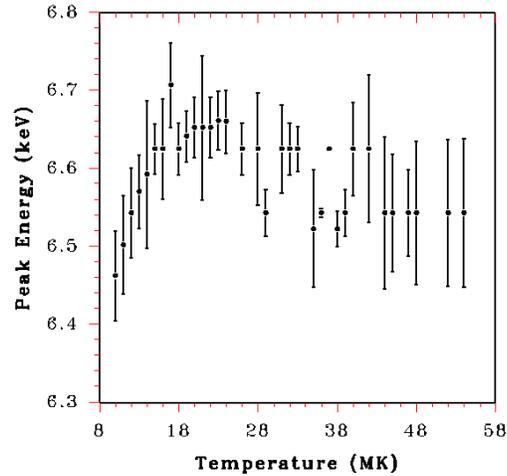

Figure 8: Variation of peak energy ($E_p$) of the Fe-line feature as a function of temperature. In Y-axis error of ± one σ in $E_p$ is shown.

3.1.3 *Equivalent Width of Fe-line features in Flare Plasma*:

The observations of the Fe line and Fe/Ni line features and neighboring continuum offer a means of determining the iron abundance $A_{flare}$(Fe) and similarly the nickel abundance $A_{flare}$(Fe/Ni) during flares by measuring their relative intensities as a function of temperature. The thermal plasma during flares is located in coronal loop structures typically $10^4$ km above the photosphere. In a chromospheric evaporation picture, this plasma is formed from the chromosphere and therefore should reflect the chromospheric composition. Fludra and Schmelz (1999) and Phillips *et al.* (2003) showed that elements with a variety of first ionization potential (FIP) are present with ratios that are characteristic of the corona *i.e.,* with low-FIP ($\leq$ 10 eV) elements enhanced by a factor of three or four but with high-FIP elements approximately the same or depleted by a factor of up to two compared with photospheric abundance. However abundance enhancements might depend upon flare intensity and duration. Thus the study of large a number of different flares is important. Fe and Ni both are low-FIP elements and therefore SLD/SOXS observations of Fe and Fe/Ni line features compared to the intensity of the nearby continuum may allow us to determine the relative abundance of Fe and Fe/Ni in flare plasma.

A measure of the Fe-line feature's intensity with respect to the continuum is provided by the equivalent width ($w$), measured in keV, defined in Equation (5) and which can be determined from Si/SLD spectra.

$$w = \int_{Line} \frac{[I(E_f) - I(E_c)]}{I(E_c)} dE \qquad (5)$$

Here $I(E_f)$ and $I(E_c)$ are the intensity of the Fe-line feature and continuum in a given channel, and $dE$ is feature width in keV.

Figure 9 shows the equivalent width (*w*) of Fe-line features as a function of temperature in six flares *viz.* 19 November 2003, 31 October 2004, 07 January 2004, 14 July 2004, 25 August 2005, and 14 August 2004 arranged in the increasing order of intensity (*cf.* Table I). It may be noted from Figure 9 that the temperature in the flares of GOES intensity <M5.0 (top panel) does not exceed 32 MK, while in higher intensity flares (bottom panel), the temperature ranges between 15 and 58 MK. Further, *w* increases exponentially with temperature up to 30 MK and then increases slowly up to 40 MK. In intense flares (bottom panel) a reduction in *w* was observed above 40 MK with increase in temperature. Our investigation suggests that the equivalent width (*w*) is independent of flare location on the Sun and rather depends upon the temperature of the flare at a given time. Thus in order to get better statistical significance we measured the equivalent width (*w*) in 135 spectra of ten M-class flares under investigation.

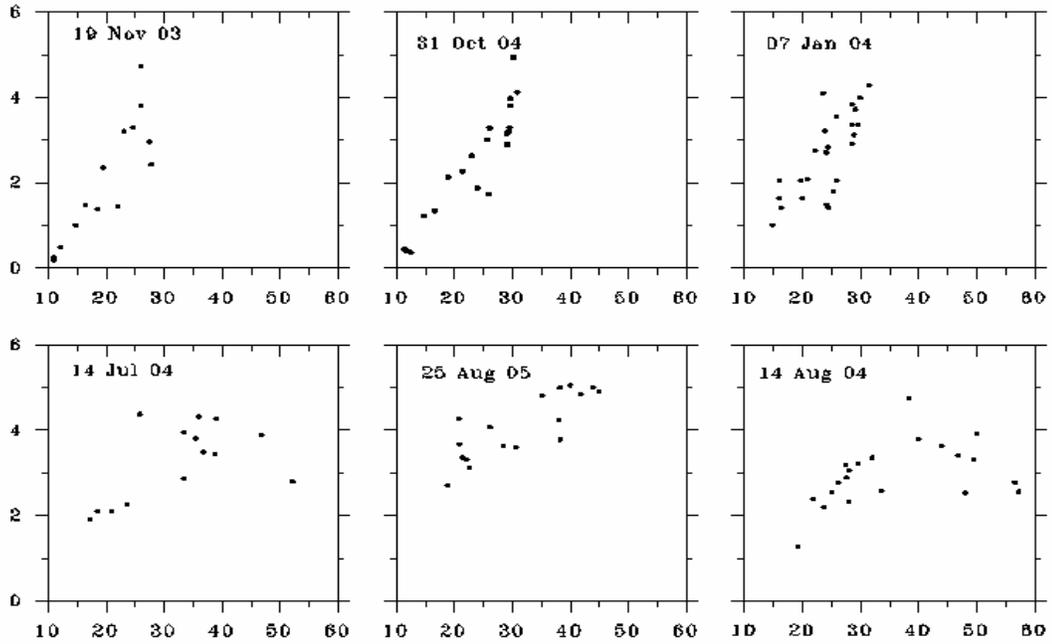

Figure 9: Variation of equivalent width (*w*) of the Fe-line feature as a function of temperature in six flares arranged in ascending order of intensity. Note that the plasma temperature in flares of intensity below M5 (top panel) does not exceed 35 MK.

Figure 10 shows the variation of *w* with temperature, which for all ten flares as in Figure 9 also shows an exponential rise of *w* until 30 MK and later rises more slowly. However it may be noted from this figure that *w* remains essentially between 3.5 and 4.0 keV in the temperature range 30 – 45 MK and then begins to fall rapidly above 45 MK. Temperatures greater than 58 MK were not found in any flare under study so the variation of *w* beyond 58 MK could not be measured.

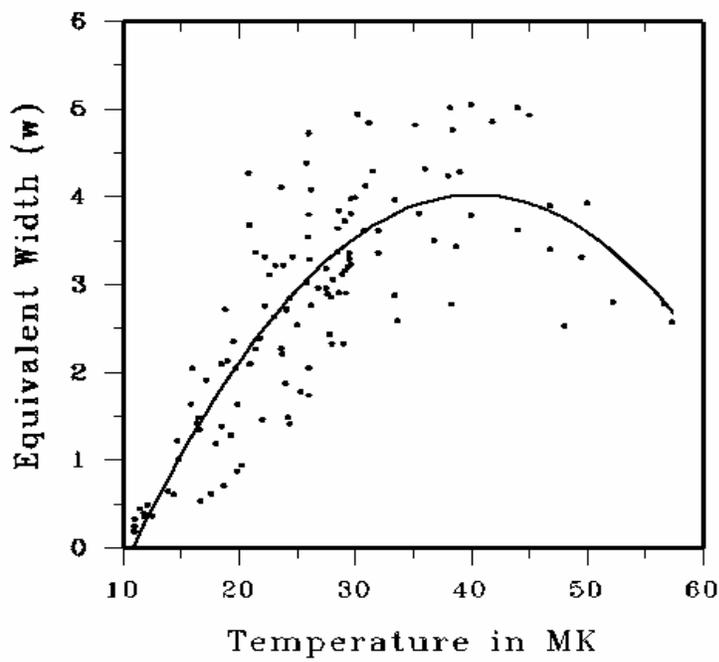

Figure 10: Variation of equivalent width (*w*) as a function of temperature combined for all ten flares under study.

## 3.2 IONIZATION STATE IN FLARES:

Phillips (2004) showed that the assumption of steady-state equilibrium in the flare plasma may not be valid due to the rapid change of temperature in the rising phase of the flare. Therefore if non-equilibrium conditions exist in the flares, the plasma would be expected to be in an ionizing state during the rising phase. However, ionization equilibrium is a good approximation unless the temperature gradient ($dT_e/dt$) ≥ 0.5 MK/s (Phillips, Neupert, and Thomas, 1974; Mewe *et al.*, 1985) or $N_e ≥ 10^{10}$ cm$^{-3}$. Thus it is very important to derive the quantity $(dT_e/dt)/T_e$ and to observe its variation over the flare duration in order to compare it with the ionization and recombination time scales. We undertook this study for four of the ten flares *viz.* 07 January 2004, 14 August 2004, 31 October 2004, and 25 August 2005 as we have long duration data for these events. We did not find ($dT_e/dt$) in excess to 0.1 MK/s, a factor five less than required for non-equilibrium conditions except in the flare of 14 August 2004 where ($dT_e/dt$) is ~0.3 MK/s and thus close to the non-equilibrium condition. However, $N_e$ is found ≥ $10^{10}$ cm$^{-3}$ throughout the flare and the temperature gradient ($dT_e/dt$) varies between -0.05 and 0.1 and therefore the flare cannot be regarded as being in equilibrium, or steady state, during the rise or decay phases. Figure 11 (a, b) shows three plots *viz.* light curves (top panel) from the Si detector in the 10 – 20 keV energy band, $T_e$ (middle panel), and $(dT_e/dt)/T_e$ (bottom panel), as a function of time. It may be noted from this figure that the intensity and temperature variations of the flares as a function of time are quite similar. However, in two cases *viz.* 31 October 2004 and 25 August 2005, during the rise phase while the intensity increases continuously temperature begins to rise only after ~100 seconds. In addition to intensity and temperature variation, the value of the parameter $(dT_e/dt)/T_e$ also varies during the rise phase as well as the decay phase.

We found that at the beginning of the flare, $(dT_e/dt)/T_e$ rises to a positive value for ~ 20 - 30 seconds and later drops to a negative value for a long period, during the decay phase. Based on the measurements of temperature gradient *i.e.,* $dT_e/dt > 0$ from our observations during rise phase of the flare, the inverse of $(dT_e/dt)/T_e$ may be considered as an ionization time, and similarly when $dT_e/dt < 0$ the inverse of $(dT_e/dt)/T_e$ may be considered as a recombination time (Phillips, 2004). We get ionization times of about 300 seconds and the whole decay phase as a recombination time but in general more than 900 seconds except in 14 August 2004 flare where ionization and recombination times are found ~90 and > 1800 seconds respectively. The ionization time scale is also representative of the ionization of $Fe^{+23}$ to $Fe^{+24}$ ions, which enables us to estimate the ionization rate coefficient ($Q_i$) as ~$10^{-12}$ (cm$^3$ s$^{-1}$). Similarly the recombination time scales represent the recombination of $Fe^{+25}$ to $Fe^{+24}$ ions allows us to estimate the recombination rate coefficient ($\alpha_i$) as ~$10^{-13}$ (cm$^3$ s$^{-1}$), assuming $N_e$ is ~$10^{10}$ cm$^{-3}$.

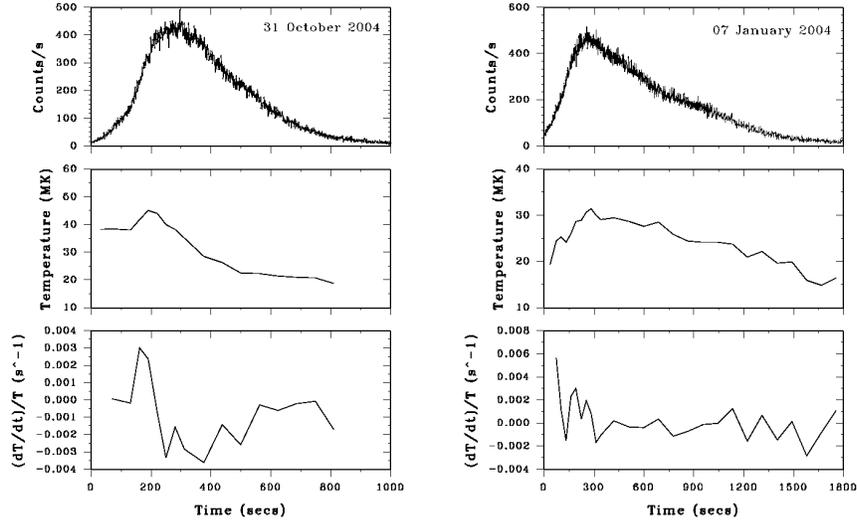

Figure 11a: Top panels: light curve of 31 October 2004 (left) and 07 January 2004 (right) flares in the 10 – 20 keV energy band as observed by Si detector of SLD/SOXS mission. Middle panels: variation of temperature ($T_e$), and, Bottom panels: variation of $(dT_e/dt)/T_e$ as a function of time in both flares.

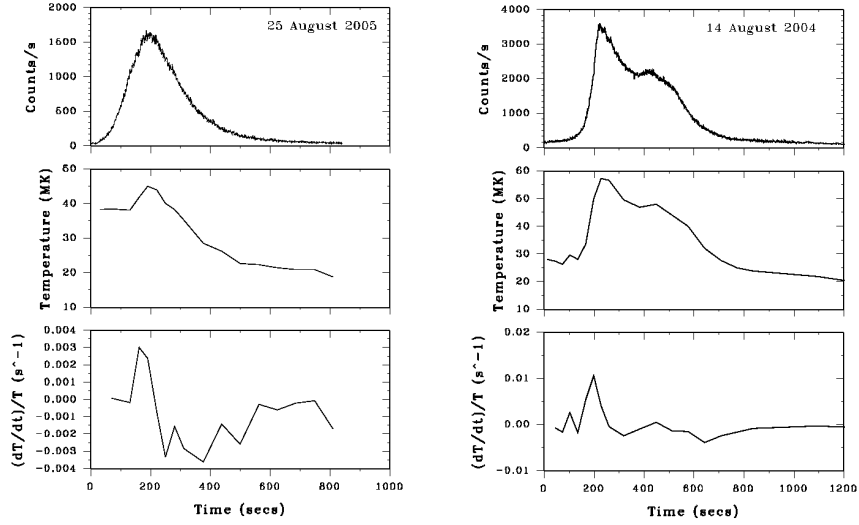

Figure 11b: Top panel: light curve of 25 August 2005 (left) and 14 August 2004 (right) flares in 10 – 20 keV energy band as observed by Si detector of SLD/SOXS mission. Middle panels: variation of temperature ($T_e$), and, Bottom panels: variation of $(dT_e/dt)/T_e$ as a function of time in both flares.

## 4. Discussion

It is well established that during the flare interval the plasma temperature varies as a function of time (Feldman *et al.,* 1995). However, in addition to this fact, our study shows that the temperature does not vary smoothly but rather fluctuates in general during the whole flare interval and particularly rapidly during the rise phase (*cf.* Figure 7 and 11). This fluctuation in flare plasma temperature ($T_e$) affects the ionization state and as a consequence we observe variation in the peak energy ($E_p$) and equivalent width (*w*) of the Fe-line emission. Our photon spectral observations from the ten flares show that the minimum critical temperature required for the Fe-line feature to be visible is 9 MK, (see Figures 8 and 9). In the temperature range 9 < $T_e$ < 30 MK the He-like Ca-line (3.86 – 3.90 keV), and He-like Fe-lines and satellites (6.4 – 6.7 keV) are most intense. Our Si detector begins spectral observations from 4 keV and it is therefore not possible to observe the Ca-line feature around 3.8 keV. However the observed increase in $T_e$ unambiguously leads to a change in peak energy because of the excitation of different principal lines of the He-like Fe XXV, satellites and resonance lines in agreement with earlier calculations (Gabriel, 1972, Boiko *et al.,* 1978, Doschek *et al.,* 1981, Feldman *et al.,* 1995, Kato *et al.,* 1997, Phillips, 2004).

The equivalent width (*w*) of the Fe-line feature is also found to vary with the $T_e$ for the flare plasma, and appears in general in agreement with those calculated by Phillips (2004) from the Chianti code using the coronal abundances of Feldman and Laming (2000). However, a detailed compassion of our measurements with the calculations of Phillips (2004) shows two important differences. . The value of *w* is significantly larger than the 3 keV predicted as a maximum by Phillips (2004) at a plasma temperature of 25 MK while the turnover temperature is 40 MK in contrast to the 25 MK calculated value. This motivated us to compare our measurements with the earlier calculations of Raymond and Smith (1977), Sarazin and Bahcall (1977), Rothenflug and Arnaud (1985), and Phillips (2004) as shown in Figure 12. It is clear from this figure that our measured values of *w* are significantly higher than earlier calculations by Raymond and Smith (1977), (referred to as RS77), Sarazin and Bahcall (1977), (referred as SB77) and Rothenflug and Arnaud (1985), (referred as RA85) in the temperature range 18 to 58 MK. However, below 18 MK our measurements are close to SB77 and RA85 although there is a steeper variation with $T_e$. On the other hand while comparing our *w* measurements with those of Philips (2004), (referred as P04), we find them significantly smaller and higher at temperature below and above 25 MK respectively.

Our findings suggest different ionic contributions to the Fe-line feature as a function of plasma temperature that governs ionization and recombination at a given time. For example the *w* value for the Fe XXV line is higher than for Fe XXIII, Fe XXIV and Fe XXVI up to 100 MK while above 100 MK emission from Fe XXVI becomes stronger and *w* is dominated by this emission. However, contributions to *w* from Fe XXII, Fe XXIII and Fe XXIV almost stop around 21, 35 and 115 MK respectively. Therefore, in the temperature range of 9 – 58 MK for the flares studied in this investigation, the major contribution to *w* may be considered to be from these ionic emissions and from Fe XXV. However, a small contribution from Fe XXVI may arise when the temperature exceeds 30 MK. Thus variation in our measured *w* as well as $E_p$ values as a function of temperature may be interpreted as due to the varying participation of different ionic emissions of Fe with temperature. Thus we may conclude that the difference in calculations of *w* by earlier investigators may be due to their selection of atomic parameters and to their choice of coronal abundance value for iron. Thus our experimental measurements of equivalent width and its variation over temperature may help to improve theoretical calculations as well as to revise the coronal abundance of Fe from that given earlier by Feldman and Laming (2000). However, in order to estimate the coronal abundance of Fe we have to measure it relative to another element. The relative abundance of Fe in the corona (Fe/H = $1.26 \times 10^{-4}$) is estimated to be four times the photospheric abundance. However, we propose to derive the relative abundance of Fe in the corona by measuring the ratio of the equivalent width (*w*) of the Fe line feature to Fe/Ni line feature as these two features are distinctly visible in our spectra. We plan to undertake this work as a future investigation.
.

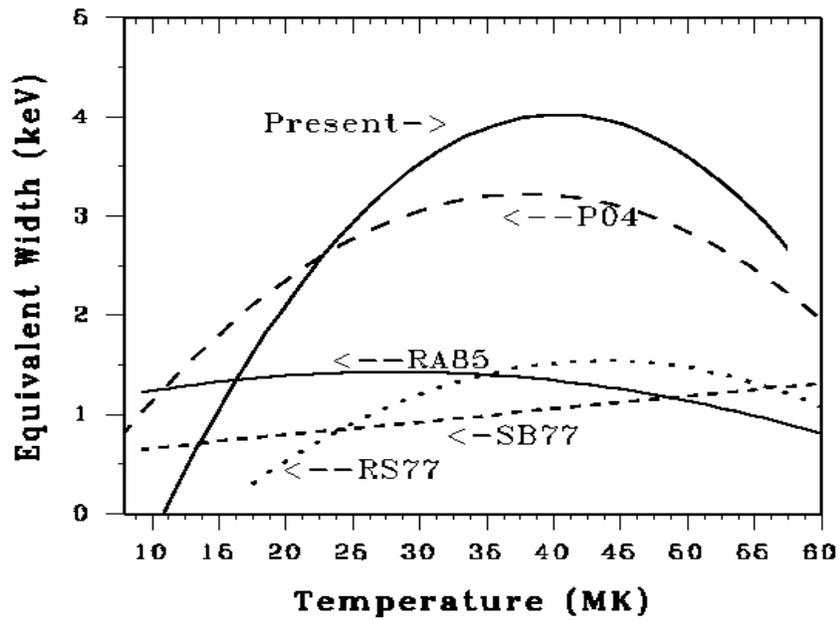

Figure 12: Comparison of our measured values of equivalent width (*w*) with previous results from RS77 (Raymond and Smith, 1977), SB77 (Sarazin and Bahcall, 1977), RA95 (Rothenflug and Arnaud, 1985), and P04 (Phillips, 2004).

## 5. Conclusion

The Si PIN detector of the SOXS Low Energy Detector (SLD) payload provides a unique opportunity to study the Fe-line and Fe/Ni-line features in great detail. In this paper we carried out a study of the Fe-line feature in order to investigate the variation of peak energy ($E_p$) and equivalent width ($w$) as a function of temperature for the flare plasma. We found that the peak energy of the Fe-line feature varies from 6.4 at 9 MK to 6.7 at 25 MK. More interestingly, equivalent width ($w$) rises exponentially up to 30 MK and then increases slowly and remains between 3.5 and 4 keV in the temperature range 30 - 40 MK. It decreases with further increase in temperature. We interpret the variation of both $E_p$ and $w$ with temperature as being due to the changes in ionization and recombination conditions in the flare plasma during the flare duration and as a consequence, the contribution from different ionic emission lines also varies. Our measurements of $w$ are compared with previous calculations and are found to be close to the results of Phillips (2004). Our measurements of $w$ may help to improve theoretical calculations of equivalent width as well as values of the relative coronal abundance of iron if compared with the intensity of the Fe/Ni-line feature.

## Acknowledgments

This investigation is result of visit of RJ to B. Dennis and K.J.H Phillips GSFC, NASA, and prolonged and very fruitful discussions. RJ is very grateful to B. Dennis for arranging this visit. We express our sincere thanks to Prof. U.R. Rao, Chairman, PRL Governing Council, and to Prof. P.C. Agrawal, TIFR, Mumbai for extensive discussions on our findings and reviewing the work. We are grateful to Prof. J.N. Goswami, Director, PRL for continuous support for research from SOXS mission. The authors express their sincere thanks to the referee for their very valuable suggestions that improved the paper.

**Figure Legends (captions)**

Figure 1: Light curves of the 31 October 2004 solar flare as recorded in L1, L2, L3, and T energy bands (see text) of Si detector of SLD/ SOXS mission

Figure 2: Count Spectra from Si PIN detector for the 31 October 2004 flare at 05:30:59 UT and pre-flare background. Note the appearance of the Fe and Fe/Ni line features during flare.

Figure 3: Photon spectra of background (left) and the flare (right) intervals on 31 October 2004. Note that photon flux in the flare is two - three orders higher than the background below 9 keV and one - two orders higher in the continuum.

Figure 4: Sequence of X-ray photon spectrum in the energy range 5 – 10 keV for the 31 October 2004 flare showing evolution of the Fe and Fe/Ni line features. The *X*-axis bar is channel width of 0.082 keV, while *Y*-axis error bar is ± one $\sigma$ of the photon flux in the given channel.

Figure 5: The X-ray photon spectra of the 14 August 2004 and 31 October 2004 at 05:43:49 UT and 05:28:59 UT respectively. Note the 9.5 – 16 keV continuum fit by isothermal plasma temperature (superhot) and emission measure. Contribution from the continuum emission down to 4 keV is shown by the extrapolated dash-dot straight line from the continuum fit. The *X*-axis bar is channel width of 0.082 keV, while *Y*-axis error bar is ± one$\sigma$ of the photon flux in the given channel.

Figure 6: The residual counts for isothermal plasma continuum fit (*cf.* Figure 6) for the 14 August 2004 and 31 October 2004 spectra..

Figure 7: Evolution of temperature and emission measure derived from the continuum of the X-ray photon spectra of the 14 August 2004 (left) and 31 October 2004 (right) flares.

Figure 8: Variation of peak energy ($E_p$) of the Fe-line feature as a function of temperature. In *Y*-axis error of ±one $\sigma$ in $E_p$ is shown.

Figure 9: Variation of equivalent width (*w*) of the Fe-line feature as a function of temperature in six flares arranged in ascending order of intensity. Note that the plasma temperature in flares of intensity below M5 (top panel) does not exceed 35 MK.

Figure 10: Variation of equivalent width (*w*) as a function of temperature combined for all ten flares under study. .

Figure 11a: Top panels: light curve of the 31 October 2004 (left) and 07 January 2004 (right) flares in 10 – 20 keV energy band as observed by Si detector of SLD/SOXS mission. Middle panels: variation of temperature ($T_e$), and, Bottom panels: variation of *($dT_e/dt)/T_e$* as a function of time in both flares.

Figure 11b: Top panel: light curve of the 25 August 2005 (left) and 14 August 2004 (right) flares in 10 – 20 keV energy band as observed by Si detector of SLD/SOXS mission. Middle panels: variation of temperature ($T_e$), and, Bottom panels: variation of $(dT_e/dt)/T_e$ as a function of time in both flares.

Figure 12: Comparison of our measured values of equivalent width (*w*) with previous results from RS77 (Raymond and Smith, 1977), SB77 (Sarazin and Bahcall, 1977), RA85 (Rothenflug and Arnaud, 1985) and P04 (Phillips, 2004).